# Photon Frequency Bears More At Less

## - Compact In-line Logic Modules For Scalable One-photon Quantum Computing Afforded By Frequency Basis

T. Asaba and S. Fukatsu*

*Graduate School of Arts and Sciences,*

*University of Tokyo at Komaba, Meguro, Tokyo 153-8902, Japan*

**Construction of an optical quantum computer (OQC)[1] is finished by implementing all necessary ingredients[2,3] with light (photon). There is, however, one more hurdle to clear. It is scalability[4], which is easily lost when accommodating many qubits by densely nesting quantum circuits. Any of the reported OQC schemes[4-7] is not necessarily best placed in this regard. Here we demonstrate the power of "frequency" degree of freedom of light, which outperforms others with its potentially infinite basis states: as multiple qubits share the same "one-photon" superposition state all along, a realistic OQC design in frequency basis adopts only one port each for input and output. As such quantum logic gates are configurable in a cascade of compact in-line modules, which ensures scalable computing. Finally, our implementation of Deutsch-Jozsa's algorithm[8] using standard laboratory laser demonstrates that frequency-basis OQC is ideally suited for such tasks even without help of nonclassicality**

**of light.**

Quantum computer (QC)[9] will find such immediate applications as large number factoring[10] and rapid database search[11]. A universal QC consists of two basic quantum logic gates (QLGs), i.e., Hadamard gate[2] and controlled-not (CNOT) gate[2,3], acting on a register of information bits called "qubits"[3]. In pursuit of appropriate basis set[5,7,12], light (photon) holds promise as many of the properties essential to QC are already equipped and conveniently accessed in experiment. Moreover, absence of direct light-light coupling warrants decoherence-free computing, which is an advantage over solid-state QCs[13]. However, from the scalability perspective, none of the known degrees of freedom of light, i.e., polarization[5], momentum[12], and photon number[7], seems to be eligible; an exponential increase of paths with increasing qubits in momentum basis compromises the scalability while polarization basis does not permit multi-qubits itself, and a QC using photon-number qubits can be a challenge[1].

Here we develop a one-photon optical QC (OQC) architecture by implementing multiple qubits utilizing frequency (standing for "angular frequency" throughout the work) degree of freedom of light (photon)[14]. With many qubits encoded in the same photon on their common path, one can design compact in-line QLG modules to be plugged in series as opposed to nesting, which allows for the scalability. These QLGs are operable upon photon frequency qubits (PFQs) created by laboratory lasers, which is exploited to demonstrate the power of one-photon OQC in implementing a quantum algorithm using frequency protocol.

A photon frequency superposition state carrying n-qubits of information, i.e., n-PFQ,

can be expanded in terms of a basis set containing $2^n$ discrete frequencies $\frac{1}{2^{n/2}}\sum_{k=0}^{2^n-1}|\omega_0 + k\Delta\omega\rangle$ where $\omega_0$ is the base frequency and $\Delta\omega$ is the frequency interval (see Supplementary Figure 1S). For example, a three-PFQ is a linear combination of $|\omega_0\rangle$ through $|\omega_0 + 7\Delta\omega\rangle$, or equivalently $|000\rangle$ through $|111\rangle$.

A Hadamard rotation acting on the second qubit maps $|000\rangle$ to $\frac{1}{\sqrt{2}}(|000\rangle + |010\rangle)$ or $\frac{1}{\sqrt{2}}(|\omega_0\rangle + |\omega_0 + 2\Delta\omega\rangle)$. Fig. 1a shows the generic frequency Hadamard gate (FHG) consisting of two cascaded asymmetric Mach-Zehnder interferometers (MZIs) where CF denotes the comb filter.

CF plays a pivotal role in our one-photon frequency OQC design. Although yet to be developed, it is essentially a "multi-chroic" edge mirror with alternate stop- and pass-bands centered at the $2^n$ frequencies. To suppress crosstalk, Rugate filter design[15] is more appropriate than stacked Bragg mirrors or a multiple splice of fiber Bragg gratings, which admit sidebands and fringes. Compared in Fig. 1b are the simulated reflectance spectra of CF1 and CF2 implemented in FHG of Fig. 1a (see Supplementary Figure 2S).

Consider entry to the FHG input of one-PFQ, i.e., a superposition state $\alpha|\omega_0,0\rangle + \beta|\omega_1,0\rangle$ where left (right) index refers to path 1 (2). CF1 discriminates frequency states and redirects them so that $|\omega_0\rangle(|\omega_1\rangle)$ takes path 1(2) whereas CF2 is used to recombine light along two paths. FSD (FSU) placed halfway down path 1 (2) represents the frequency shifter which downconverts $|\omega_1\rangle$ to $|\omega_0\rangle$ (upconverts $|\omega_0\rangle$ to $|\omega_1\rangle$). To put either $\{\alpha = 1, \beta = 0\}$ or $\{\alpha = 0, \beta = 1\}$ in the output $\frac{1}{\sqrt{2}}\{(\alpha - \beta)|\omega_0,0\rangle - (\alpha + \beta)|\omega_1,0\rangle\}$ yields a Hadamard transform without loss, i.e., unitary. Note that FHG generates multi-qubits, i.e.,

n-PFQ, simply by inserting FHG n-times.

A quantum phase gate (QPG) is a variant of conditional two-qubit gate[3] equivalent to a CNOT gate[16]. FQPG is conditioned on the status of control qubit that is built in the n-PFQ itself as prepared and taken along with the rest of (data) qubits. This allows FQPG to act upon n-PFQ on its path: there is no extra path just for the control qubit unlike conventional QPG. The generic frequency QPG (FQPG) is illustrated in Fig. 1c. PS is the phase shifter that flips the sign of all qubits on path 2 at once so that one-PFQ $\alpha|\omega_0,0\rangle+\beta|\omega_1,0\rangle$ is mapped to $\alpha|\omega_0,0\rangle-\beta|\omega_1,0\rangle$ at the output. Thus FQPG operates at a 100-% throughput, i.e., unitary.

In experiment, the following points are noted. First, the frequency QLGs are designed to act upon one-photon states, with their operation and detection building upon first-order interference as described later. This makes their implementation easier as coherent states delivered from standard laboratory lasers are eligible, which holds a competitive advantage over the QC schemes where the nonclassical nature such as bipartite entanglement[17] arguably plays a major role. Second, a special read-out technique must be adopted in frequency basis as time-varying amplitudes that occur due to mixing of light with unlike frequencies smear out stationary fringes over an ensemble. We herewith attempt direct capture of heterodyne beats, i.e., one-photon interference fringes in time domain. Lastly, we must be able to demonstrate frequency QLGs and OQC even without CFs.

Figure 2a shows a realistic experimental setup to emulate FHG implementation without CFs, which contrasts with the model case shown in Fig. 1a though the essence remains the same. A superposition state of identical frequency with a relative phase shift $\theta$ is prepared within first MZI equipped with PS, i.e., a wedge prism on a translation stage providing a phase lag along path 1. It is then sent to the two input ports of second MZI where two frequency shifters, FS1 and FS2, driven at the frequencies at a small detuning, $\delta = |\omega_1 - \omega_2|$, should generate one-photon beats at the output. The normalized light intensity at the detector for input of 1 is

$$\frac{1}{4}(1 + \sin\delta t \cos\theta)\cdot \qquad (1)$$

Thus the Hadamard rotation is manifest as modulation of amplitude as $\theta$ varies. The prefactor 1/4 accounts for the loss due to the choice of half mirror (NPBS) instead of CF at the exit. Figure 2b shows the as-captured beat traces as a function of $\theta$. There is essentially no shift in the phase of heterodyne beats as $\theta$ is varied. The beat amplitude is plotted in Fig. 2c versus displacement proportional to $\theta$ where the solid line is a fit using Equation (1). The near perfect match clearly supports that FHG operates as theory predicts.

Next, the implementation of FQPG was attempted. Once FQPG is available, a composite CNOT gate is ready to be constructed by putting FQPG in between two FHGs in series with a bypass loop (see Supplementary Figure 3S). Similarly to FHG, the experimental circuit design of Fig. 3a was modified from Fig. 1c to emulate FQPG without CFs. FS1 and FS2 in first MZI cause beats at a frequency detuning $\delta = |\omega_1 - \omega_2|$. The idea here is that entanglement allows FQPG to function even without CFs, i.e., the only point the nonclassical properties of light enter. We place a half-wave plate (HWP1) on the same arm as FS1 (path1), which rotates the horizontal polarization of light by 90 degrees. The superposition state after first MZI is thereby tagged with polarization, and hence gets hyper-entangled[18] with path (momentum), frequency and polarization such that

$$\frac{i}{\sqrt{2}}|\omega_0,0\rangle_H+\frac{i}{\sqrt{2}}|\omega_1,0\rangle_V. \tag{2}$$

The frequency states are now distinguishable by referring only to polarization, which enables us to design polarization-entangled FQPG without CFs. We used polarization beam splitters (PBSs) are used to separate and combine light of orthogonal polarizations.

A second HWP (HWP2) placed after PS on path 2 of second MZI is a disentangler, which flips vertical polarization back to horizontal. The light intensity at the detector reads

$$\frac{1}{4}\{1+\cos(\delta t+\phi)\}. \tag{3}$$

Thus FQPG operation simply changes the phase of the heterodyne beats. The prefactor 1/4 appears again as before. Figure 3b shows the beats recorded as a function of relative phase shift $\phi$ controlled by variable delay as light passes through a quartz wedge prism (lower panel of Fig. 3c). As visible in the upper panel of Fig. 3c, the beat amplitude is leveled off regardless of wedge displacement and hence $\phi$ values. This is what Eq. 3 predicts and taken as evidence for successful implementation of FQPG.

Finally, one-qubit frequency OQC was constructed to implement the Deutsch-Jozsa's algorithm[8] upon one-photon state. Two FHGs are arranged in series because FQPG is not of importancee here (Fig. 4a). A pair of FSUs is driven here at null detuning $\delta=0$ to produce stationary fringes along path 2. The two quantum black boxes[8], BB1 and BB2, made of wedge prisms provide fixed phase lags in preset combinations. As captured in Fig. 4b, fringes and anti-fringes develop for like (constant) settings {BB1, BB2}= {0,0} and {1,1} and unlike (balanced) settings {BB1, BB2}= {0,1} and {1,0}, respectively, where "1" indicates "$\pi$-delayed" while "0" means "no phase lag". This successful demonstration clearly points to the usefulness of classical light in implementing frequency OQC.

In terms of scalability[1,2], frequency OQC holds an advantage that cascaded optical circuits take only one path. This is assured by the directional and convergent properties pertinent to CFs, as opposed to momentum basis where the scalability is lowered each time a bifurcating-optics doubles the path. However, we face a trade-off that demanding a large number of frequency basis states in turn requires as many combs. The design and fabrication of CFs would be of complexity and even labor-intensive. Nevertheless, compact in-line modular design of frequency QLGs, will help reduce component count in building one-photon frequency OQCs.

More importantly, frequency degree of freedom of light, if combined with momentum (path) or photon number, can create multi-qubits in greater number ever, thereby boosting the potential of "one-photon" OQC.

## Methods

We used a He-Ne laser (632.8 nm) as the light source. The intensity autocorrelation measurement indicates that outputs are of coherent states or Poissonian, i.e., a signature of classical light[19] (see Supplemenatry Figure 4S). We maintained the single-path design in setting up experiment. We used largely linear optics besides acousto-optic modulators driven near 80 MHz for use as FSs despite their smaller-than-ideal conversion efficiency ≈85 %. Precision balanced frequency shift reversal was made possible by sharing a common driving source (see Supplemenatry Figure 5S). For data capture and read-outs, an uncooled photon-number-resolving avalanche photodiode-based solid-state detector with a 10-MHz bandwidth was used. This was direct-current coupled to an oscilloscope and transient readings were captured as one-photon heterodyne beats in the low MHz range.

## Acknowledgments

The authors acknowledge Y. Yasutake for his technical assistance in preparation of measurement setup.

## Additional information

The authors declare herewith that there are no competing financial interests. All correspondence and requests for materials should be addressed to S.F.

Figures and figure captions

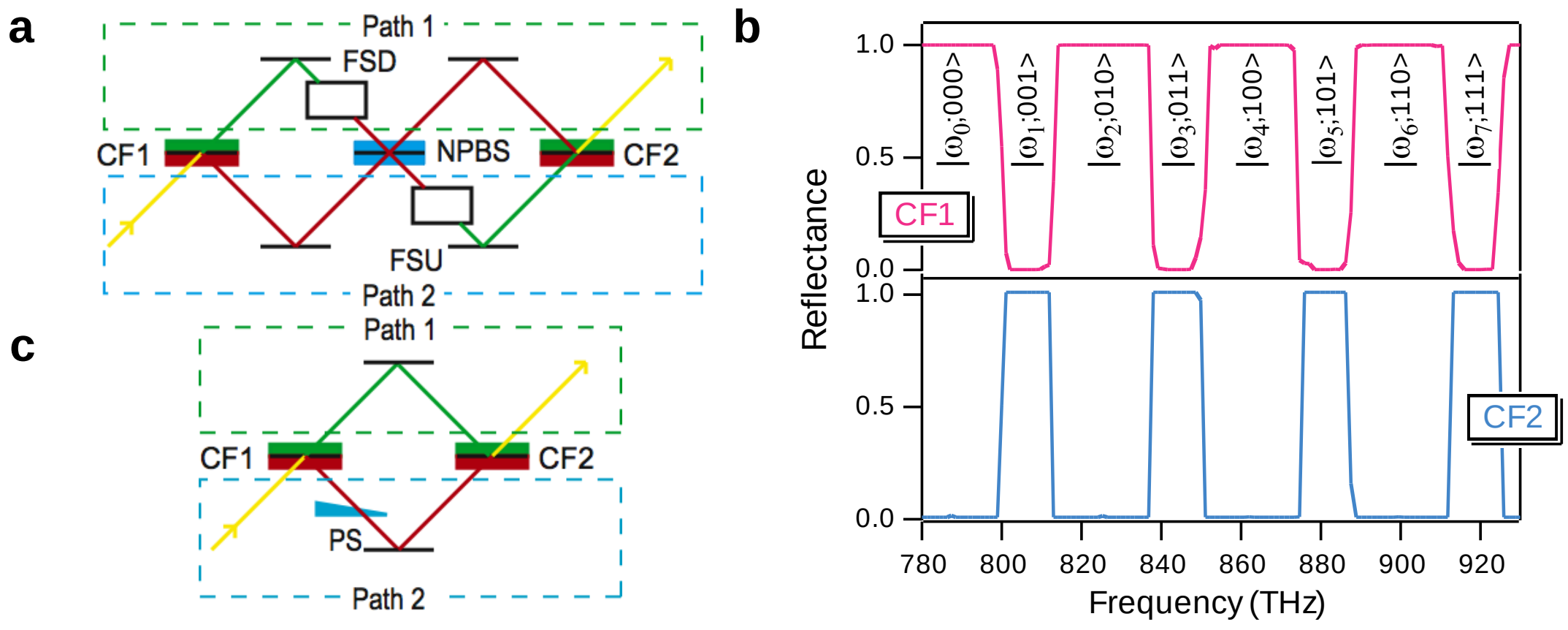

**Figure 1  Optical quantum circuits for frequency-QLGs and simulated characteristics of CFs. a,** Schematic FHG. CF1(2) separates (recombines) light at the input (output); FSD(U), frequency shifter, which downconverts (upconverts) the frequency of light as it is transmitted; NPBS, nonpolarizing beam splitter. **b,** Reflectance spectra of typical three-qubit CFs. **c,** Schematic FQPG. PS, phase shifter.

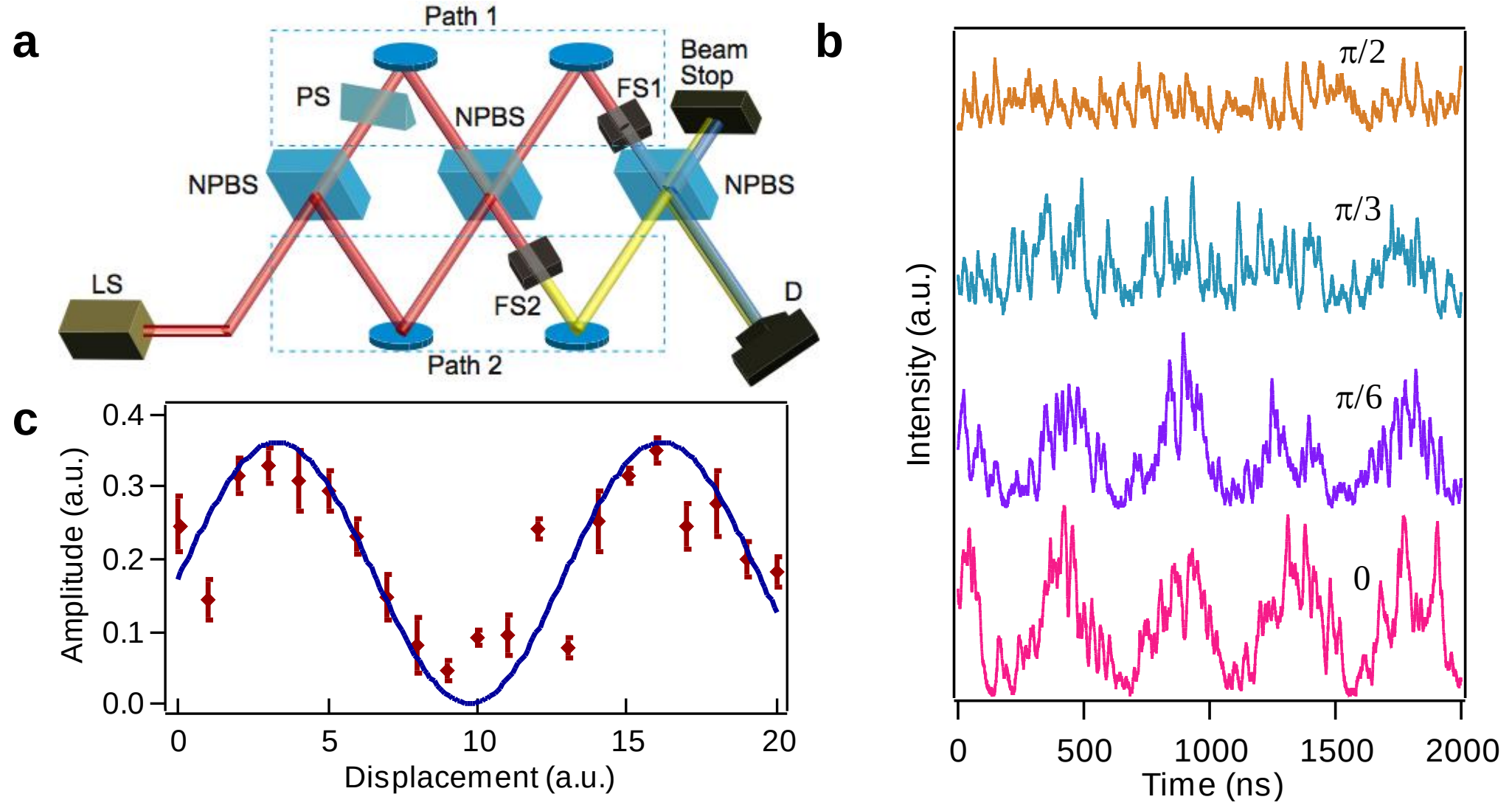

**Figure 2  Characterization of Hadamard rotation acting upon one-PFQ without CFs. a,** Schematic FHG. LS, light source; PS, phase shifter, which provides an input phase lag on path 1; NPBS, nonpolarizing beam splitter; FS1, FS2, frequency shifter; D, detector. **b,** Oscilloscope traces of as-captured beats as a function of $\theta$. Note the absence of phase shift. **c,** Beat amplitude versus $\theta$. Solid line is a theoretical fit using Eq. 1.

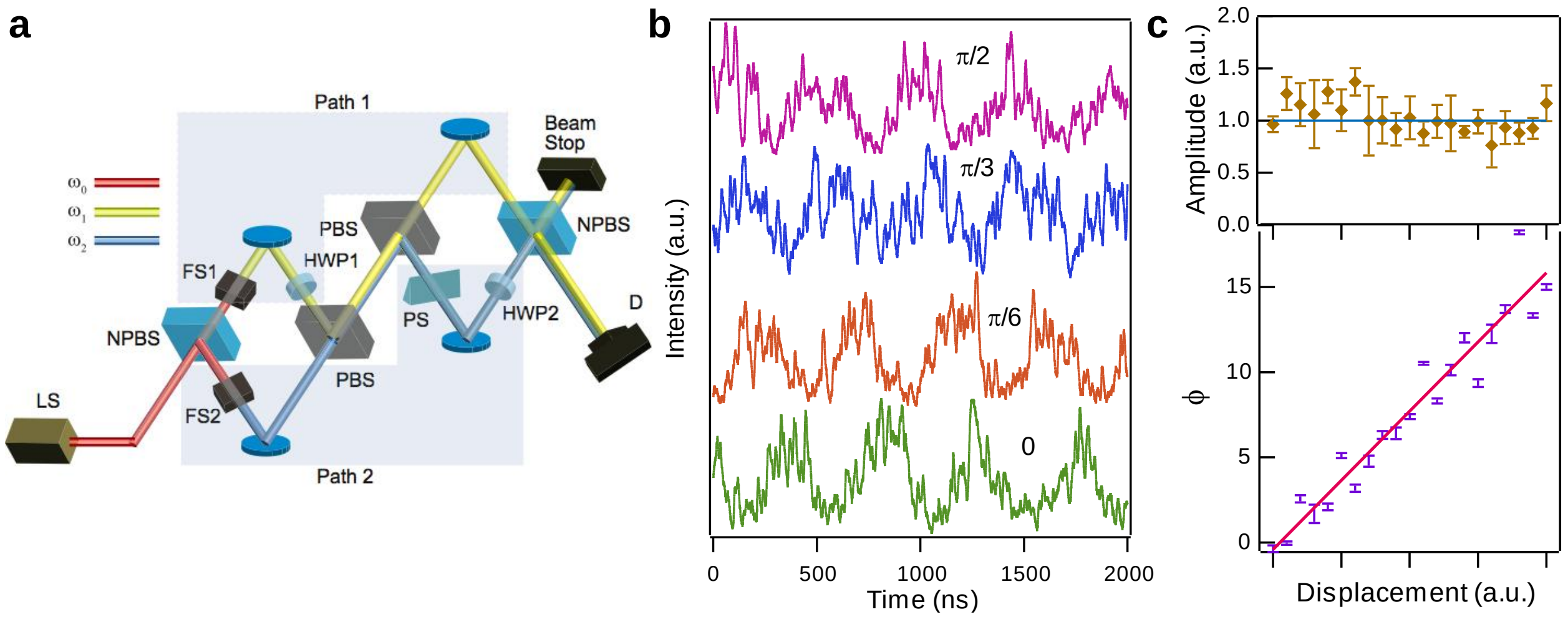

**Figure 3 Characterization of polarization-entangled FQPG experiment performed on one-PFQ without CFs. a,** Schematic one-qubit FQPG. First MZI creates frequency-path (momentum)–polarization-entangled one-PFQ. LS, light source; NPBS, nonpolarizing beam splitter; HWP1(2), half-wave plate as entangler (disentangler), which rotates light polarization by 90 degrees; PBS, polarizing beam splitter; FS1, FS2, frequency shifter; PS, phase shifter; D, detector. **b,** Oscilloscope traces of captured beats as a function of phase shift $\phi$. **c,** Beat amplitude (upper) and $\phi$ (lower) versus wedge prism displacement. Solid lines are to guide the eye.

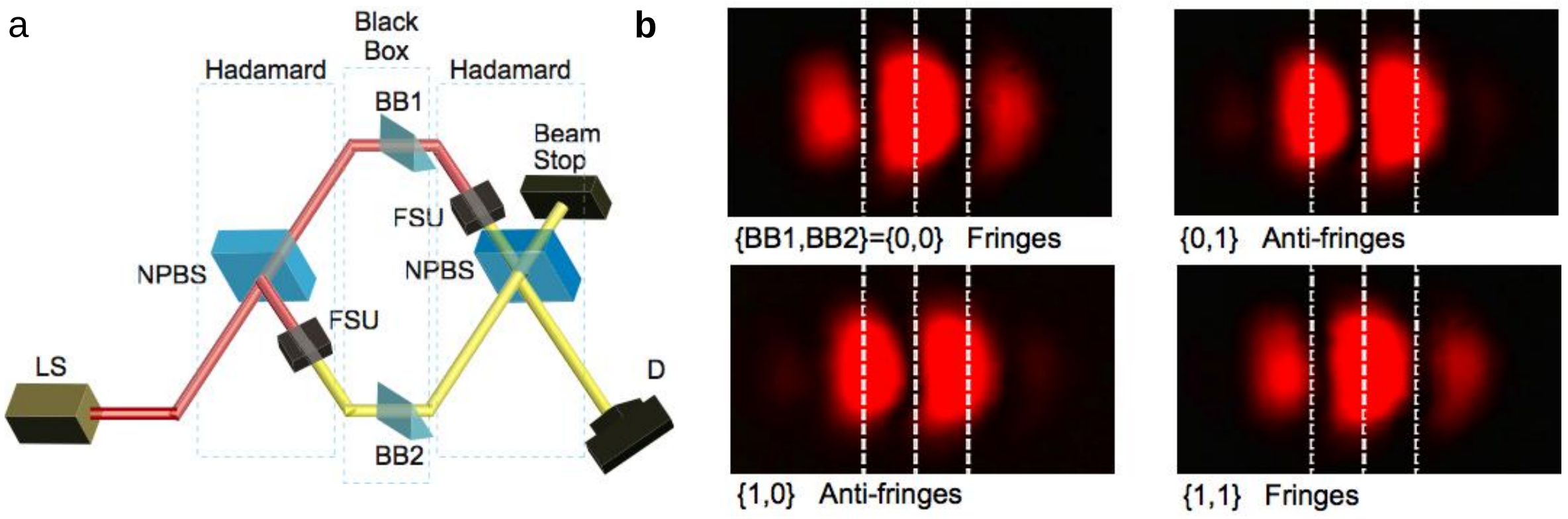

**Figure 4 Frequency-basis implementation of Deutsch-Jozsa's algorithm using one-photon inteference. a,** Schematic one-qubit photon frequency QC made of two cascaded FHGs. LS, light source; NPBS: nopolarizing beam splitter; FSU, frequency upconverters; BB1, BB2, quantum black box providing preprogrammed delays; D, detector. **b,** Far-field patterns: fringes and anti-fringes appear for like settings {BB1, BB2}= {0,0} and {1,1} and unlike settings {BB1, BB2}= {0,1} and {1,0}, respectively, where "1" indicates "$\pi$-delayed".